\documentclass[twocolumn,preprintnumbers,aps,prd,superscriptaddress,nofootinbib,tightenlines]{revtex4-1}
\pdfoutput=1

\usepackage{amsmath,amssymb,amsbsy,amstext, amsthm, simplewick}  
\usepackage{graphicx}   
\usepackage{verbatim}   
\usepackage{color}      
\usepackage{subfigure}  
\usepackage{hyperref}   

\usepackage{enumerate}
\usepackage{empheq}
\usepackage{booktabs}
\usepackage{amsfonts}
\usepackage{wasysym}
\usepackage{wrapfig}
\usepackage{multirow}
\usepackage{array}

\usepackage{colortbl}
\definecolor{summersky}{cmyk}{0.71,0.33,0,0.5}
\definecolor{flamingo}{cmyk}{0,0.51,0.71,0.5}
\definecolor{rp}{cmyk}{0.2, 1, 0.6, 0}
\definecolor{pacificblue}{cmyk}{0.95,0.3,0, 0.5}
\definecolor{gray60}{cmyk}{0.4,0.4,0,0.8}
\graphicspath{
    {Figures/}
}

\newcommand{\be}{\begin{eqnarray} }
\newcommand{\ee}{\end{eqnarray} }
\newcommand{\bs}{\begin{split} }
\newcommand{\es}{\end{split} }

\newcommand{\la}{\langle}

\newcommand{\ra}{\rangle}

\newcommand{\nc}{\newcommand}

\newcommand{\bvec}[1]{\mathbf{#1}}
\nc{\nn}{\nonumber}
\nc{\eps}{\epsilon}

\begin{document}

\title{The Boostless Bootstrap and BCFW Momentum Shifts}


\author{David Stefanyszyn}
\email{d.stefanyszyn@damtp.cam.ac.uk}
\author{Jakub Supe\l{} }
\email{js2154@cam.ac.uk}

\affiliation{Department of Applied Mathematics and Theoretical Physics, Centre for Mathematical Sciences,
University of Cambridge, Wilberforce Road, Cambridge CB3 0WA, UK}

\begin{abstract}
\noindent  
In a recent paper \cite{PSS}, three-particle interactions without invariance under Lorentz boosts were constrained by demanding that they yield tree-level four-particle scattering amplitudes with singularities as dictated by unitarity and locality. In this brief note, we show how to obtain an independent verification and consistency check of these \textit{boostless bootstrap} results using BCFW momentum shifts. We point out that the constructibility criterion, related to the behaviour of the deformed amplitude at infinite BCFW parameter $z$, is not strictly necessary to obtain non-trivial constraints for the three-particle interactions.
\end{abstract}

\maketitle

\section{Introduction}
The on-shell approach to computing scattering amplitudes had led to tremendous advances in our understanding of gauge theory and gravity \cite{Elvang:2013cua,TASI,Benincasa:2013}. Weinberg's seminal papers from the 60's \cite{Weinberg:1964ew,Weinberg:1965rz} showed that the combination of Poincar\'{e} invariance and unitarity uniquely picks out Maxwell's equations as the description of a massless spin-$1$ particle (photon) and Einstein's equations as the description of a massless spin-$2$ particle (graviton), at tree-level. In addition, charge conservation and the equivalence principle follow from consistency of scattering amplitudes involving photons and gravitons, respectively. \\


Since then, there have been many attempts to constrain or \textit{bootstrap} interactions using a purely on-shell approach, by invoking the principles of unitarity, locality and causality. One of these attempts is the technique known as BCFW deformations \cite{BCFW}, in which some of the external momenta are deformed by a complex parameter $z$. Consideration of the analytic structure of the amplitude as a function of $z$ often imposes non-trivial constraints on interactions \cite{Benincasa:2007xk}, as we will explain in detail in Section \ref{sec:SH}. Constraints can also be derived in another way \cite{NAH,McGadyRodina,Schuster:2008nh}, without deforming the amplitude, but rather by demanding consistent factorisation. Here one simply writes down the most general form of the tree-level four-particle amplitude, and demands that poles only arise due to intermediate (exchanged) particles going on-shell, with the corresponding residues given by a product of on-shell three-particle amplitudes. \\

Notable results of this bootstrap programme include the observations that long-range forces cannot be mediated by particles with spin $\geq 3$, Yang-Mills (YM) is the unique theory of multiple massless spin-$1$ particles, General Relativity (GR) is the unique low-energy theory of a massless spin-$2$ particle, and the existence of massless spin-$3/2$ particles requires gravity and supersymmetry leading uniquely to supergravity\footnote{One can also show that there is an upper limit on the number of spin-$3/2$ particles that a unitary theory can contain \cite{McGadyRodina}.} \cite{Benincasa:2007xk,McGadyRodina}.  This on-shell approach shows that the low-energy properties of Poincar\'{e} invariant theories are virtually inevitable.  \\

Given such spectacular success, attention has recently turned to applying similar methods to spatial correlation functions which are the fundamental observables in cosmology, see \cite{Arkani-Hamed:2018kmz,Baumann:2019oyu,Arkani-Hamed:2015bza,Sleight:2019mgd,Sleight:2019hfp,Baumann:2020dch,Green:2020ebl,Mata:2012bx,Ghosh:2014kba,Kundu:2014gxa,Arkani-Hamed:2017fdk,Arkani-Hamed:2018bjr,Benincasa:2018ssx,Benincasa:2019vqr,Hillman:2019wgh,Benincasa:2020aoj,COT} and references therein. These cosmological correlators live on the late-time boundary of an approximate de Sitter spacetime and encode details about the bulk spacetime through their dependence on spatial momenta. A key ingredient in this cosmological bootstrap is the fact that correlators contain flat-space scattering amplitudes in the residues of their total-energy poles \cite{Maldacena:2011nz,Raju:2012zr} (for a detailed proof see \cite{COT}). These amplitudes form part of the theoretical data required to bootstrap the corresponding correlators, and therefore the cosmological bootstrap programme requires us to have a solid understanding of flat-space scattering amplitudes. \\

Much effort has so far focused on correlators fixed by de Sitter symmetries (or conformal symmetries on the boundary). Intuitively, taking the cosmological bootstrap programme beyond exact de Sitter symmetries in order to construct inflationary correlators, requires taking the $S$-matrix bootstrap beyond exact Poincar\'{e} symmetries. Motivated by this, a recent paper \cite{PSS} derived the singular parts of tree-level four-particle amplitudes where the free propagators are assumed to be massless and Poincar\'{e} invariant, i.e. all dispersion relations are linear with each particle propagating at the same speed, but with interactions that are allowed to break Lorentz boosts. The breaking of Lorentz boosts was assumed to enter only through time derivatives, with the fields being Lorentz-covariant tensors. Within this set-up, analytically continued three-particle amplitudes are fixed by the helicities of the external particles up to an almost arbitrary function of their energies and these functions are then constrained by demanding consistent factorisation, but without making use of BCFW shifts. The most interesting results from this \textit{boostless bootstrap} are \cite{PSS}:
\begin{itemize}
\item When there is a graviton in the spectrum, all three-particle interactions must reduce to their Poincar\'{e} invariant form, even those that do not involve the graviton itself\footnote{Related results based on field-theoretic methods have been derived in \cite{Emergent1,Emergent2,Emergent3}. See also \cite{SoftGravitons} where Lorentz boosts arise in the presence of soft gravitons.}. Universal coupling of gravity to all other particles is then recovered. 
\item Low-energy self-interactions of a photon must vanish. The leading allowed operator has six derivatives and is therefore mass dimension-$9$. 
\item There exists at least one large class of boost-breaking theories involving scalars, spin-$1/2$ fermions and photons (\textit{boost-breaking massless QED}). The corresponding Lagrangians contain generalised, boost-breaking gauge redundancies. 
\end{itemize}  

In the Poincar\'{e} invariant cases, all the results derived by imposing consistent factorisation have also been reached via the BCFW formalism. Likewise, in this paper we show that the results of \cite{PSS} can be derived using BCFW shifts as a tool to automate consistent factorisation and therefore provide a neat consistency check. In Section \ref{sec:SH} we briefly introduce the spinor helicity formalism that we will use throughout, present general three-particle amplitudes for boost-breaking theories, and briefly review BCFW momentum shifts. In Section \ref{sec:Constructibility} we argue that BCFW shifts remain a useful tool for non-constructible theories by making a distinction between \textit{accessible} and \textit{inaccessible} singularities. In this case, while the four-particle amplitudes are not completely fixed by the three-particle ones, we can still use BCFW shifts to constrain the latter. We illustrate this in Section \ref{sec:4p} for boost-breaking theories of self-interacting spin-$S$ particles. We end the paper with some concluding remarks. \\

Before moving to the main body of this paper, let us emphasise that in \cite{PSS} a certain form of the four-particle amplitudes was assumed. The assumption was that the helicity scaling of the external particles is given by angle and square brackets only and not from off-diagonal round brackets. This is a restricted form of the four-particle amplitudes which nevertheless captures a wide class of theories. This assumption amounts to assuming that the underlying Lagrangian is constructed out of Lorentz covariant fields with the breaking of Lorentz boosts driven by the freedom to add time derivatives at will. In a theory with $SO(3)$ covariant fields, we expect some off-diagonal round brackets too and in \cite{PSS} this was shown to be the case for a particular Framid amplitude where the Framid constitutes the Goldstone modes of spontaneously broken Lorentz boosts \cite{Zoology}. We will follow \cite{PSS} and assume that the helicity scaling of four-particle amplitudes is fixed by angle and square brackets only. Extending our results to more general amplitudes is an interesting avenue for future work. 

\section{Spinor helicity formalism and BCFW deformations} \label{sec:SH}

We work in four spacetime dimensions and use the spinor helicity formalism to present amplitudes in a compact form. In this formalism, a complex null four-momentum $p^{\mu}$ is represented as a product of two-component spinors as\footnote{The inverse of this equation is $p^{\mu} = \frac{1}{2}(\bar{\sigma}^{\mu})^{\dot{\alpha}\alpha} p_{\alpha \dot{\alpha}}$ and so the energy of a particle in terms of the spinors is $E = \frac{1}{2}(\bar{\sigma}^{0})^{\dot{\alpha}\alpha} \lambda_{\alpha} \tilde{\lambda}_{\dot{\alpha}}$.}
\begin{equation}
p_{\alpha \dot{\alpha}} = \sigma^{\mu}_{\alpha \dot{\alpha}}p_{\mu} = \lambda_{\alpha} \tilde{\lambda}_{\dot{\alpha}},
\end{equation}
where $\sigma^{\mu}_{\alpha \dot{\alpha}}$ are the Pauli matrices and the undotted and dotted indices transform in the spinor representations of the Lorentz group i.e. $(1/2,0)$ and $(0,1/2)$ respectively. Throughout we follow the conventions of \cite{Dreiner} and we assume that each particle satisfies $p^{\mu}p_{\mu} = E^2 - \bvec{p}^2= 0$ on-shell.
When boosts are broken, amplitudes are constructed from $SO(3)$ invariant quantities rather than $SO(1,3)$ invariant ones. In \cite{PSS}, it was shown that such three-particle amplitudes are functions of the following objects:
\begin{itemize}
\item ``angle'' brackets: $\la i  j \ra  =  \epsilon^{\alpha \beta} \lambda^{(i)}_{\alpha} \lambda^{(j)}_{\beta}$,
\item  ``square'' brackets: $[ i  j ]  =  \epsilon^{\dot{\alpha} \dot{\beta}} \tilde{\lambda}^{(i)}_{\dot{\alpha}} \tilde{\lambda}^{(j)}_{\dot{\beta}}$,
\item  energies: $E_i$.
\end{itemize}
Here latin indices label the external particles: $i,j = 1,2,3$. We remind the reader that the spinors are \textit{not} grassmanian and therefore the angle and square brackets are anti-symmetric. By demanding that the amplitudes scale in the appropriate way under helicity transformations, on-shell, non-perturbative three-particle amplitudes take the form \cite{PSS}
\begin{equation} 
\mathcal{A}_3 = \begin{cases}
\la 1 2 \ra^{ d_3} \la 2 3 \ra^{d_1} \la 3 1 \ra^{d_2}F^H(E_1, E_2, E_3),  &h <  0,  \\
[ 1 2 ]^{ -d_3} [ 2 3 ]^{-d_1} [ 3 1 ]^{-d_2}F^{AH}(E_1, E_2, E_3) , & h > 0,
\end{cases}
\label{eq:3pFinalResult}
\end{equation}
where $d_i = 2 h_i - h$, $h_{i}$ is the helicity of the $i^{\text{th}}$ particle and $h$ is the sum of the helicities (if $h=0$, the amplitude can be a sum of the two expressions.) The presence of functions\footnote{The superscripts refer to holomorphic and anti-holomorphic kinematic configurations \cite{Benincasa:2007xk}.} $F^H, F^{AH}$ (which depend on the helicities) reflects the fact that Lorentz boosts are no longer an assumed symmetry. The Poincar\'{e} invariant amplitudes are recovered when these functions are constant. As energy is conserved, we will often consider the $F$'s as functions of two arguments only. As an example, the three-particle amplitude for three incoming gravitons, two with negative helicity and one with positive helicity, is 
\begin{align}
\mathcal{A}_{3}(1^{-2},2^{-2},3^{+2}) = \left(\frac{\la 12 \ra^3}{\la 23 \ra \la 31 \ra}\right)^{2}F^{H}(E_{1},E_{2}).
\end{align}
In this case, Bose symmetry dictates that $F^{H}(E_{1},E_{2}) = F^{H}(E_{2},E_{1})$. \\

Now, to be consistent with unitarity and locality, a tree-level four-particle amplitude must factorise into a product of on-shell three-particle amplitudes on each of its poles. Poles correspond to exchanged particles going on-shell and consistent factorisation dictates that, for example,
\begin{equation}
\lim_{s \to 0} \left(s \mathcal{A}_4 \right) = \mathcal{A}_3(1,2, -I) \times \mathcal{A}_3(3,4, I),
\end{equation}
where $s  = (p_1 + p_2)^2$ is the propagator of the exchanged particle which is labelled by $I$. Analogous relations hold when $t = (p_1 + p_3)^2 \to 0$ and $u =  (p_1 + p_4)^2 \to 0$. For future reference, in terms of the spinors these Mandelstam variables are given by $s = \la 12 \ra [12] = \la 34 \ra [34]$, $t = \la 13 \ra [13] = \la 24 \ra [24]$ and $u = \la 14 \ra [14] = \la 23 \ra [23]$. Requiring the amplitude to factorise correctly on each pole is often highly non-trivial \cite{PSS,Benincasa:2007xk,NAH,McGadyRodina,Schuster:2008nh} since, for example, the $s$-channel residue can contain poles in $t$ and $u$ which then need to be interpreted as propagation of a particle in those channels. This is beautifully illustrated for multiple spin-$1$ particles where consistency in each channel requires the coupling constants to satisfy the Jacobi identity. \\

In \cite{Benincasa:2007xk}, Benincasa and Cachazo elegantly used BCFW shifts \cite{BCFW} to formally assess this consistency for a number of tree-level four-particle amplitudes. The simplest BCFW shift takes two particles $i$ and $j$ and deforms their energies and momenta according to
\begin{equation}
\lambda^{(i)}(z) = \lambda^{(i)} + z \lambda^{(j)}, \quad \tilde{\lambda}^{(j)}(z) = \tilde{\lambda}^{(j)} - z \tilde{\lambda}^{(i)},
\end{equation}
with all other spinors kept fixed. This choice preserves the on-shell conditions for both particles as well as energy-momentum conservation. The deformed amplitude $\mathcal{A}_{4}^{(i,j)}(z)$ is a rational function of the complex parameter $z$ and thus can be fully deduced from knowledge of its poles, residues and behaviour at infinity. It takes the form
\begin{equation}
\mathcal{A}_{4}^{(i,j)}(z) = \sum\limits_n \frac{res_{z = z_n} \mathcal{A}_{4}^{(i,j)}(z)  }{z-z_n} + B^{(i,j)}(z),
\end{equation}
where the boundary term $B^{(i,j)}(z)$ is regular in the entire complex plane. For four-particle amplitudes, only two poles can be reached by a given deformation (since $p_{i} + p_{j}$ is independent of $z$) and as we remarked above, the corresponding residues are evaluated from the three-particle amplitudes alone. For example, summing over the possible helicities of the exchanged particle, for $\mathcal{A}_{4}^{(1,2)}(z)$ the two poles that can be reached are $z_{t}$ and $z_{u}$ and we have
\begin{align} \label{Deformation12}
\mathcal{A}_{4}^{(1,2)}(z) = & \sum\limits_{h_I} \frac{A_3(\hat{1},3,-\hat{I}) A_3(\hat{2},4, \hat{I}) }{t(z)} \nonumber \\
+ &   \sum\limits_{h_I} \frac{A_3(\hat{1},4,-\hat{I}) A_3(\hat{2},3, \hat{I}) }{u(z)} \nonumber  \\
+ & B^{(1,2)}(z),
\end{align}
where a hat indicates that the particle has had one of its spinors deformed and evaluated at the appropriate pole. We have $t(z) = (p_{1}(z) + p_{3})^2 = \la 13 \ra [13] + z \la 23 \ra [13]$ and $u(z) = (p_{1}(z) + p_{4})^2 = \la 14 \ra [14] + z \la 24 \ra [14]$ and therefore the locations of the poles are
\begin{align}
z_{t} = -\frac{\la 13 \ra}{\la 23 \ra}, \qquad z_{u} =- \frac{\la 14 \ra}{\la 24 \ra}.
\end{align}
$\mathcal{A}_{4}^{(i,j)}(0)$ corresponds to the amplitude for unshifted momenta, and then constraints on three-particle couplings can be derived by demanding that distinct $\mathcal{A}_{4}^{(i,j)}(0)$ coincide at $z = 0$ \cite{Benincasa:2007xk} i.e.
\begin{equation}
\mathcal{A}_{4}^{(i,j)}(0) = \mathcal{A}_{4}^{(k,l)}(0) \quad \forall \quad i,j,k,l.
\label{eq:MijMkl}
\end{equation}
This is the \textit{four-particle test}.
\section{Constructibility criterion} \label{sec:Constructibility}
In its original formulation, the above described method is reserved for \textit{constructible} theories for which $B^{(i,j)}(z)$ vanishes. In this case the singular parts of undeformed amplitudes can be compared with one another and the full four-particle amplitude is determined by the three-particle ones. Since $B^{(i,j)}(z)$ is regular, it is sufficient to prove that the amplitude tends to zero as $z \to \infty$. This is usually a non-trivial matter, necessitating a reference to the Lagrangian and a detailed counting of powers of momenta. Fortunately, many theories describing nature are constructible including, most notably, YM \cite{BCF,BCFW} and GR \cite{Benincasa:2007qj} (scalar field theories are not constructible in the sense described above. This has lead to new, interesting momentum shifts and on-shell recursion relations being derived for scalar theories with non-linearly realised symmetries \cite{ScalarRecursion1,ScalarRecursion2}). \\ 

However, for the boost-breaking amplitudes of interest here, it is unlikely that $B^{(i,j)}(z)$ would vanish, since the unknown functions of energies will in general contribute positive powers of $z$ to the tree-level amplitude. Indeed, for both particles $i$ and $j$, the deformation of their energies is linear in $z$ and so the divergence at large $z$ gets worse as additional powers of energy are included:
\begin{align}
E_i(z) & = E_i(0) + \frac{z}{2}(\bar{\sigma}^{0})^{\dot{\alpha}\alpha} \lambda^{(j)}_{\alpha} \tilde{\lambda}^{(i)}_{\dot{\alpha}}, \\
E_j(z) & = E_j(0) - \frac{z}{2}(\bar{\sigma}^{0})^{\dot{\alpha}\alpha} \lambda^{(j)}_{\alpha} \tilde{\lambda}^{(i)}_{\dot{\alpha}}.
\end{align}

The BCFW method can still be useful for non-constructible theories, however. One possibility is to introduce a distinction between \textit{accessible} and \textit{inaccessible} singularities of $\mathcal{A}_{4}^{(i,j)}(z)$, along the lines of \cite{Feng:2014pia}. We say a singularity is \textit{accessible} via a deformation of momenta $i$ and $j$ if this singularity is approached as $z \to z_*$ for some $z_*$. Otherwise we say it is \textit{inaccessible}. The regular term $B^{(i,j)}(z)$, by definition, cannot have any singularities in the $z-$plane and therefore cannot contribute to any residues of the accessible singularities of $\mathcal{A}_{4}^{(i,j)}(z)$. But it may exhibit inaccessible singularities. As an illustration of this distinction, consider a single scalar theory which is famously non-constructible. In the absence of additional global charges, the three-particle amplitude is a non-zero constant, $\mathcal{A}_3 = g$, and so we have
\begin{eqnarray}
\mathcal{A}_{4}^{(1,2)}(0) &=& g^2 \left(\frac{1}{t} + \frac{1}{u}\right) + B^{(1,2)}(0), \\
\mathcal{A}_{4}^{(1,4)}(0) &=& g^2 \left(\frac{1}{s} + \frac{1}{t}\right) + B^{(1,4)}(0).
\end{eqnarray}
The consistency condition $\mathcal{A}_{4}^{(1,2)}(0)  = \mathcal{A}_{4}^{(1,4)}(0) $ can be satisfied by choosing $ B^{(1,2)}(z) = \frac{g^2}{s}$ and $B^{(1,4)}(z) = \frac{g^2}{u}$, since these two functions do not have any accessible singularities with regards to their own deformations. \\

In the following section we will constrain boost-breaking amplitudes using the fact that the regular term $B^{(i,j)}(z)$ does not have any accessible singularities. We will see that for spinning particles, we can derive the highly non-trivial constraints first found in \cite{PSS}. 

\section{Constraining three-particle interactions} \label{sec:4p}
In this section we will constrain three-particle interactions for theories of a single spin-$S$ particle with integer $S$. We will derive the constraints first presented in \cite{PSS}. We also checked that the BCFW techniques allow us to recover other results in \cite{PSS}, namely those of (gravitational) Compton scattering and the full analysis for a scalar or a photon coupled to gravity. Those calculations contain only minor differences compared with what is presented below so we omit the details in favour of brevity. We remind the reader that we do not impose boost invariance, but only demand that the free theory is Poincar\'{e} invariant, with the on-shell condition $E^2 - \bvec{p}^2 = 0$ for each particle, and that boost violations enter the action only through time derivatives. \\

Consider the amplitude $\mathcal{A}_4(1^{+S}2^{-S}3^{+S}4^{-S})$, where superscripts denote the helicities of incoming particles of some integer spin $S$. We will impose matching conditions between deformations $(1,2)$ and $(1,4)$. First consider $(1,2)$. Using the expressions for\footnote{These amplitudes arise from the leading order couplings. Higher-dimension operators give rise to the $A_{3}(1^{+S},2^{+S},3^{+S})$, $A_{3}(1^{-S},2^{-S},3^{-S})$ amplitudes but we don't consider these here. We refer the reader to \cite{PSS} for a discussion on these amplitudes.}  $A_{3}(1^{+S},2^{+S},3^{-S})$ and $A_{3}(1^{-S},2^{-S},3^{+S})$ given in \eqref{eq:3pFinalResult}, and\footnote{Here we have assumed $[13]$ and $[14]$ are non-zero, and therefore $t=0$ and $u=0$ are approached as $\la 13 \ra = 0$ and $\la 14 \ra = 0$ respectively (or as $[24]=0$ and $[23]=0$ respectively, by momentum conservation).}
\begin{eqnarray}
p_{1}(z_t)+p_{3} & = & \frac{[13]}{[14]} \lambda^{(3)} \tilde{\lambda}^{(4)}, \\
p_{1}(z_u)+p_{4} & = & \frac{[14]}{[13]} \lambda^{(4)} \tilde{\lambda}^{(3)}, \label{IntermediateU}
\end{eqnarray}
to eliminate all copies of $\lambda^{(I)}$ and $\tilde{\lambda}^{(I)}$, which are the spinors associated with the exchanged particle\footnote{For example, in the $t$-channel we set $\lambda^{(I)} = \alpha \lambda^{(3)}$ and $\tilde{\lambda}^{(I)} = \beta \tilde{\lambda}^{4}$ with $\alpha \beta = \frac{[13]}{[14]}$. When computing the residue, $\alpha$ and $\beta$ only appear in the product $\alpha \beta$.}, we find 
 \begin{eqnarray}
&& \mathcal{A}_{4}^{(1,2)}(0) =  B^{(1,2)}(0) + \nonumber \\
&&  \left( \frac{1}{t} F_{\hat{1},3}F_{\hat{2},4} + \frac{1}{u}  F_{\hat{1},-\hat{1}-4}F_{\hat{2},-\hat{2}-3}\right) \left(\frac{ [13]^2 \la 24 \ra^2}{s} \right)^S.
\end{eqnarray}
In the $u$-channel we have summed over the two possibilities for the helicity configuration of the exchanged particle but given \eqref{IntermediateU}, only one of these is non-zero.
To keep formulae compact, here we have introduced subscripts to the $F$'s to denote their arguments e.g. $F(E_{i},E_{j}) \equiv F_{i,j}$ and $F(E_{i},E_{j}+E_{k}) \equiv F_{i,j+k}$. Again, hats denote deformed objects evaluated at the appropriate points e.g. in the $1/t$ coefficient, $F_{\hat{1},3} \equiv F(\hat{E}_{1}(z_t),E_{3})$ where $\hat{E}_{1}(z_t)$ is the deformed energy of particle $1$ evaluated at $z = z_t$. Likewise, in the $1/u$ coefficient, hatted energies are evaluated at $z=z_{u}$. We have also removed the $H/AH$ superscripts since the functions are identical, due to parity, up to an inconsequential overall sign \cite{PSS}. \\

Now, we can also write 
 \begin{eqnarray} \label{M12Final}
&& \mathcal{A}_{4}^{(1,2)}(0) =  \tilde{B}^{(1,2)}(0) + \nonumber \\
&&  \left( \frac{1}{t} F_{1,3}F_{2,4} + \frac{1}{u}  F_{1,-1-4}F_{2,-2-3}\right) \left(\frac{ [13]^2 \la 24 \ra^2}{s} \right)^S,
\end{eqnarray}
where here we dropped the hat above all the energies, which indicates that the expression is evaluated at their \textit{undeformed} values. This can be justified as follows. We assume that the $F$'s can be Taylor expanded around the undeformed energies. The deformed energies are
\begin{align}
&\hat{E}_{1}(z_{t}) = E_{1} - \frac{t}{2 \la 23 \ra [13]} (\bar{\sigma}^0)^{\dot{\alpha} \alpha}\lambda^{(2)}_{\alpha}\tilde{\lambda}^{(1)}_{ \dot{\alpha}}, \\
&\hat{E}_{2}(z_{t}) = E_{2} + \frac{t}{2 \la 23 \ra [13]} (\bar{\sigma}^0)^{\dot{\alpha} \alpha}\lambda^{(2)}_{\alpha}\tilde{\lambda}^{(1)}_{ \dot{\alpha}},
\end{align} 
with similar expressions evaluated at $z = z_u$. For the class of Lagrangians considered in this paper, energies appear in $F_{a,b}$ only with non-negative powers, and each factor of an energy is generated by a time derivative acting on the field. In the above, we see that potential new singularities generated by the deformed energies are all \textit{inaccessible}, as they correspond to the vanishing of $\la 23 \ra $ or $[13]$, but these do not depend on $z$. Moreover, only the leading term in the Taylor expansion will exhibit accessible singularities, since in all subleading terms $t$ and $u$ will be cancelled out. We can therefore simply absorb all subleading terms into $B^{(1,2)}$, thus introducing $\tilde{B}^{(1,2)}$ that still does not contain any terms singular in $t$ or $u$. Although it could become singular in some kinematic configurations, especially at $s=0$, that is not a problem, because this singularity is \textit{inaccessible} and we only demand that $\tilde{B}^{(1,2)}$, for those configurations for which it can be defined, does not have any singularities as a function of $z$. \\
%

We now play the same game for the $(1,4)$ deformation which amounts to interchanging particles $2$ and $4$. We have
\begin{eqnarray} \label{M14Final}
&& \mathcal{A}_{4}^{(1,4)}(0) = \tilde{B}^{(1,4)}(0) +  \nonumber \\
&& \left( \frac{1}{t} F_{1,3}F_{4,2} + \frac{1}{s}  F_{1,-1-2} F_{4,-4-3} \right) \left(\frac{ [13]^2 \la 24 \ra^2}{u} \right)^S.
\end{eqnarray}
We discussed the $S=0$ case earlier where we showed that equating $\mathcal{A}_{4}^{(1,2)}(0)$ and $\mathcal{A}_{4}^{(1,4)}(0)$ requires us to make certain choices for the boundary terms. Let us now consider $S >0$ with $S$ an integer. We see that $\mathcal{A}_{4}^{(1,2)}(0)$ in \eqref{M12Final} contains terms proportional to $1/(ts^S)$ and $1/(us^S)$ which are both singular in more than one Mandelstam variable and thus cannot be accounted for or modified by $\tilde{B}^{(1,2)}(0)$ nor $\tilde{B}^{(1,4)}(0)$. A similar observation applies to $\mathcal{A}_{4}^{(1,4)}(0)$ in \eqref{M14Final}.
Thus, by matching the amplitudes we find the necessary condition
\begin{equation}
\frac{a}{s^{S}t} + \frac{b}{s^{S}u} = \frac{c}{u^{S}t} + \frac{d}{u^{S}s},
\end{equation}
where
\begin{eqnarray}
a & = &  F_{1,3}F_{2,4}, \\
b & = &  F_{1,-1-4} F_{2,-2-3}, \\
c & = & F_{1,3}F_{4,2}, \\
d & = & F_{1,-1-2}F_{4,-4-3}.
\end{eqnarray}
Recalling that $s+t+u = 0$, this constraint, given that it must be valid for all kinematics, is equivalent to
\begin{align}
a u^{S} - b (s+u) u^{S-1} - c s^{S} + d (s+u) s^{S-1} =0.
\end{align}
For $S=1$ we therefore have $a = (b-d) = - c$, or equivalently, 
\begin{align}
F_{1,3} F_{2,4} & - F_{1,-1-4}F_{2,-2-3} + F_{1,-1-2} F_{4,-4-3} = 0,
\end{align}
which is simply an alternative form of (4.25) from \cite{PSS}. Assuming that the $F$'s are polynomials, in \cite{PSS} it was shown that the only solution to this system is $F \equiv 0$ once we impose that the $S=1$ functions are alternating polynomials as dictated by Bose symmetry\footnote{The spinor helicity parts of the $S=1$ three-particle amplitudes are odd under the exchange of identical particles so if the amplitudes are to be even by Bose symmetry, the $F$'s must be alternating.}. We therefore see that the leading order three-particle interactions for three-photons must vanish, as is the case for Poincar\'{e} invariant theories. For $S=2$ we require $a=b =c=d$, or equivalently, 
\begin{align}
F_{1,3}F_{2,4} = F_{1,-1-4} F_{2,-2-3} = F_{1,-1-2}F_{4,-4-3},
\end{align}
which gives rise to the constraints $(4.31)-(4.33)$ from \cite{PSS} once we use the fact that the $S=2$ functions are symmetric in their arguments by Bose symmetry\footnote{For $S=2$, the spinor parts of the three-particle amplitudes are even under the exchange of identical particles and so the $F$'s are symmetric polynomials.} (this also makes the $a=c$ constraint trival). In \cite{PSS} it was shown that the only solution to this system is $F=constant$ and so again the three-particle interactions are reduced to their Poincar\'{e} invariant form, but this time the amplitudes are non-zero and are those of GR. Finally, for $S > 2$, it is simple to see that $a=b=c=d=0$ is required and therefore there are no consistent three-particle interactions for these massless, higher-spin particles even when boosts are broken, as was also concluded in \cite{PSS}. 

\section{Summary}\label{sec:Discussion}

Very recently, the singular parts of four-particle amplitudes were bootstrapped in \cite{PSS} by demanding that they factorise into a product of on-shell three-particle amplitudes on simple poles. In that work, consistent factorisation was implemented directly without making use of BCFW momentum shifts. In this short note, we have shown that the same results can be derived by using BCFW shifts to automate consistent factorisation. We presented full details for the illustrative cases of single spin-$S$ particle amplitudes but have also checked that the procedure produces the expected results for Compton scattering, and its gravitational analogue, as well as for scalars or photons coupled minimally to gravity. For single spin-$S$ particles, the boostless bootstrap teaches us that the leading three-particle couplings for a photon must vanish, the leading three-particle couplings for a graviton must be those of GR, while massless higher-spinning particles do not self-interact. For photon Compton scattering, boost-breaking interactions between the photon, scalars and spin-$1/2$ fermions are allowed and can be described by Lagrangians with generalised boost-breaking gauge redundancies. For gravitational Compton scattering, all couplings must reduce to their boost-invariant counter-parts with universal couplings of all particles to gravity. Finally, scalars and photons that are minimally coupled to gravity are forced to have Poincar\'{e} invariant self-interactions (constant or vanishing, respectively for the scalar and the photon). For full details see \cite{PSS}.  \\

Although the theories we have considered are not \textit{a priori} constructible, in the sense that the boundary terms do not necessarily vanish at large $z$, we have still been able to use BCFW shifts to constrain the three-particle couplings contributing to particle exchange. This does mean, of course, that the three-particle amplitudes themselves do not fully fix the four-particle ones. Indeed, all of the four-particle amplitudes we have constructed are defined up to the presence of ``contact" terms that are regular for all kinematic configurations. It would be very interesting to investigate the possibility of using generalised momentum shifts, possibly along the lines of \cite{Raju2}, to recursively derive exact higher-point amplitudes even if only for a subset of boost-breaking theories. It would also be very interesting to investigate the generalised on-shell recursion relations introduced in \cite{Benincasa:2011kn}, where boundary terms are fixed with additional knowledge of a subset of the zeros of the deformed amplitude, in our boost-breaking setting.

\section*{Acknowledgements}
We would like to thank Paolo Benincasa, Tanguy Grall, Sadra Jazayeri and Enrico Pajer for useful discussions and comments on a draft of this paper. D.S. has been supported in part by the research program VIDI with Project No. 680-47-535, which is (partly) financed by the Netherlands Organisation for Scientific Research (NWO). J.S. has been supported by a grant from STFC.

\bibliographystyle{apsrev4-1}
\bibliography{refs}

\end{document}